\documentclass[11pt]{article}%
\usepackage{amsfonts}
\usepackage{graphicx}
\usepackage{amsmath}
\usepackage{amssymb}%
\usepackage{tikz-cd}
\usepackage[applemac]{inputenc}
\usepackage[cmtip,all]{xy}

\newcommand{\proj}[1]{\ensuremath{\mathbb{P}^{#1}}}

\pdfoutput=1
\makeatletter
\renewcommand{\theequation}{\thesection.\arabic{equation}}
\@addtoreset{equation}{section}
\makeatother
\renewcommand{\a}{\alpha}
\renewcommand{\b}{\beta}

\newcommand*{\defeq}{\mathrel{\vcenter{\baselineskip0.5ex \lineskiplimit0pt
                     \hbox{\scriptsize.}\hbox{\scriptsize.}}}%
                     =}

\newcommand{\beq}{\begin{equation}}
\newcommand{\eeq}{\end{equation}}
\newcommand{\bear}{\begin{eqnarray}}
\newcommand{\eear}{\end{eqnarray}}

\newcommand{\be}{\begin{eqnarray}}
\newcommand{\ee}{\end{eqnarray}}

\begin{document}
\baselineskip=18pt
\baselineskip 0.65cm

\begin{titlepage}
\setcounter{page}{0}
\renewcommand{\thefootnote}{\fnsymbol{footnote}}
\begin{flushright}
ARC-18-14\\
\end{flushright}
\vskip 1.5cm
\begin{center}
{\Huge \bf
$A_\infty$-Algebra from Supermanifolds
}

\vspace{1cm}
{\large
Roberto Catenacci$^{~a,c,d}$\footnote{roberto.catenacci@uniupo.it},
Pietro Antonio Grassi, $^{~a,b,d,}$\footnote{pietro.grassi@uniupo.it}
and
Simone Noja
$^{~d,e,f,}$\footnote{simone.noja@uninsubria.it}

\medskip
}
\vskip 0.5cm
{
\small
\centerline{$^{(a)}$ \it Dipartimento di Scienze e Innovazione Tecnologica,}
\centerline{\it Universit\`a del Piemonte Orientale,}
\centerline{\it viale T. Michel, 11, 15121 Alessandria, Italy}
\medskip
\centerline{$^{(b)}$ \it INFN, Sezione di Torino,}
\centerline{\it via P. Giuria 1, 10125 Torino, Italy}
\medskip
\centerline{$^{(c)}$ \it Gruppo Nazionale di Fisica Matematica, }
\centerline{\it INdAM, P.le Aldo Moro 5, 00185 Roma, Italy} 
\medskip
\centerline{$^{(d)}$ \it Arnold Regge Center,}
\centerline{\it via P. Giuria , 1, 10125 Torino, Italy}
\medskip
\centerline{$^{(e)}$ \it Dipartimento di Scienze e Alta Tecnologia (DiSAT),}
\centerline{\it Universit\`a degli Studi dell'Insubria,}
\centerline{\it via Valleggio 11, 22100 Como, Italy}
\medskip
\centerline{$^{(f)}$ \it INFN, Sezione di Milano}
\centerline{\it via G. Celoria 16, 20133 Milano, Italy} 
}
\vskip  .3cm
\medskip
\end{center}
\smallskip
\centerline{{\bf Abstract}} 
\medskip
\noindent
{Inspired by the analogy between different types of differential forms  
on supermanifolds and string fields in superstring theory, 
we construct new multilinear non-associative products of forms 
which yield an $A_\infty$-algebra. 
}
\vskip  .5cm

\noindent
\end{titlepage}
\setcounter{page}{1}


\vfill
\eject
\newpage\setcounter{footnote}{0} \newpage\setcounter{footnote}{0}

\section{Introduction}

Quite recently the discovery of some new algebraic and geometric structures in physics and mathematics has led to a renewed interest in the study of supermanifolds and their peculiar geometry \cite{ CNR, CCGhd, CCGGeom, CDGM, Catenacci:2018xsv,DonWit, PiGeo, Witten}.   
One of the main motivation comes from superstring theory. As is well-know there are two ways to 
construct the supersymmetric sigma model representing the perturbative 
Lagrangian of superstrings: 1) the Ramond-Nevue-Schwarz (RNS) sigma model, with world-sheet anti-commuting spinors and world-sheet supersymmetry and 2) the Green-Schwarz/Pure Spinor 
(GS/PS) sigma model, with anti-commuting target space spinors and target space supersymmetry. 

The RNS formulation has several interesting pros, but difficulties appear when one computes higher genus amplitudes. The quantization procedure by gauge-fixing the world-sheet supergravity leads to ghost 
insertions in the  conformal field theory correlation functions for anomaly cancellation. 
Nonetheless, some of those insertions are rather delicate and they lead to inconsistencies at higher loops. 
These insertions are the so-called \emph{Picture Changing Operators} ($Z$ or $X$ and $Y$) introduced in \cite{FMS,GWS}. 
Recently, string perturbation theory has been revised by E. Witten \cite{WittenPerturbation}, pointing 
out that some of the inconsistencies can be by-passed by a suitable integration theory of forms 
on supermanifolds and, in particular, on super Riemann surfaces. As reviewed in \cite{VorGeom, Witten}, the integration theory of forms on supermanifolds takes into account the complex of \emph{integral forms}. The latter are distributional-like forms, usually written as $\delta(d\theta)$, which serve to control the integration over the $d\theta$'s. The need of integral forms in the context of superstrings and super Riemann surfaces was already pointed out by Belopolski \cite{Belo1} and discussed, from a strictly mathematical point of view, by Manin \cite{Manin} and other authors. In particular, in \cite{Belo1} a practical way to handle forms was discussed and described. 

On the other hand, for GS/PS sigma model, it was discovered \cite{Berkovits:2004px, Berkovits:2004dt,Berkovits:2006vi} that amplitude computations (at tree level and at higher orders) also require the insertion of Picture Changing Operators to cancel the anomalies and to make the amplitudes meaningful. In this context, 
the geometry related to those operators was less clear, so that they were built in analogy with RNS superstrings. 
Nonetheless, eventually, they appear to be integral forms for the target superspace (we recall that in GS/PS sigma model, 
the quantum fields are maps from a given Riemann surface to a target supermanifold) and for them the usual 
rule of Cartan differential calculus can be used.  

Both the case of PCO's for RNS and GS/PS superstrings can be understood from a pure geometrical point of view by insisting on having a meaningful integration theory for supermanifolds (in the RSN case, integration on  super Riemann surfaces with given punctures and boundaries, while in the case of GS/PS for target space supermanifolds with a given supermetric), which in turn requires to understand at a deeper level the peculiarities arising whenever part of the geometry is anticommuting.  
In particular, it turns out that a new number - beside the ordinary form degree - is needed to describe forms on a supermanifold \cite{Witten}: this is called {\it picture number} and, essentially, it counts the number 
of delta functions of the differential $1$-forms, namely the $\delta(d\theta_i)$. As discussed in several 
papers (see for example \cite{CDGM}), two Dirac delta forms anti-commute and that implies an upper bound to the 
number of delta forms that can appear in a given form. 
In particular, the picture number can range from zero (in which case we denote 
those forms as {\it superforms}) to the maximum value which coincides with the fermionic dimension of the 
supermanifold. In that case we refer to the related complex as to the {\it integral forms} complex (see also \cite{Witten}): in particular, working on a supermanifold of dimension $n|p$, forms of degree $n$ and picture $p$ are actually sections of the Berezinian sheaf, and they can be integrated over. If the fermionic dimension of the superamanifold is greater than one, then between superforms - having picture number equal to zero - and integral forms - having picture number equal to the fermionic dimension of the supermanifold -, we can have forms having a middle-dimensional picture, that are not superforms, nor integral forms, namely they have some Dirac delta functions, but less than the maximum possible number. These are called {\it pseudoforms}. So far, for the sake of exposition only an algebraic characterization of superforms, integral forms and pseudoforms has been hinted. Notheless, remarkably, a sheaf theoretical description, disclosing interesting relations and dualities, might be given  
 \cite{Catenacci:2018xsv}. 

Once that the ``zoo'' of forms is established, differential operators relating these forms can be defined. 
Besides the usual differential operators $d$, it emerges a new differential operator denoted by $\eta$ 
which anticommutes with $d$ and it is nilpotent. This is physically motivated by the embedding of the $N=1$ RNS superstring 
into a $N=4$ supersymmetric sigma model as shown in \cite{Berkovits:1994vy}.  In the language of $N=4$ superconformal symmetry the two operators $d$ and $\eta$ 
are the two anticommuting supercharges of the superconformal algebra. 
Furthermore, $\eta$ is crucial for a useful characterization of the superstring Hilbert space. 
Indeed, that Hilbert space contains those states generated by quantizing the superghosts 
 (Small Hilbert Space (SHS) \cite{FMS,pol}) The same Hilbert space can be represented in terms of a different set of quantized fields and two descriptions are identical by excluding the zero mode of one of these quantum fields (to be precise the superghosts $\beta$ and $\gamma$ are reparametrized by two fermionic degrees of freedom and one bosonic degree of freedom). On the other hand, including a zero mode, the Hilbert space gets doubled, leading 
 to the so-called Large Hilbert Space (LHS). The original SHS lives in the kernel of $\eta$, but in LHS 
 additional structures emerge leading to an explicit solution of the constraints \cite{Berkovits:1994vy,Erler}. 
 
Translating this set up in geometric terms on supermanifolds, we found that 
the LHS corresponds to an enlarged set of forms which contains also {\it inverse forms} (see \cite{Catenacci:2018xsv}) 
on which the corresponding operator $\eta$ can be built. Similarly as above, excluding inverse forms is achieved by imposing $\eta$-closure on these extended complexes. 
Again, drawing from string theory experience, we can construct two additional operators known as PCO $Z$ and $Y$. They are built in terms of Dirac delta function integral representation and they act on the entire space 
of forms \cite{Belo1,Catenacci:2018xsv}. 

The graded (supersymmetric) wedge product makes forms on a supermanifold in an algebra. In particular given two forms on a supermanifold, the wedge product acts by adding their form degrees - as it is usual -, and also their picture numbers. It follows that, in general, it maps two forms into a forms having a greater (or equal, in a limiting case) picture number. 
 The operators $d$ and $\eta$ act as derivations on the exterior algebra of forms, while the PCO's are \emph{not} derivation with respect to the wedge product. \\
String theory, in particular its second quantized version, \emph{superstring field theory}, has actually yet another construction to hint \cite{Erler}. Indeed, over the years, there have been several proposed actions reproducing the full fledged superstring spectrum where the insertion of 
PCO's is crucial (see for instance \cite{Witten:1986qs,Preitschopf:1989fc} and see \cite{Erler} for 
further examples). Nonetheless, none of them turned out to be fully consistent. One of the main problem is due to the location of PCO's insertions. For first quantized amplitudes, the position of the PCO's is harmless, since on-shell the insertion turns out to be position-independent. On the contrary, for an off-shell second quantized action that it is pivotal. The position of the PCO's breaks the gauge invariance of the theory, leading to inconsistent results. To avoid this problem, new multilinear operations forming a \emph{non-associative algebra} known as $A_\infty$-algebra have been proposed recently by Erler, Konopka and Sachs in \cite{Erler}.

Again, mimicking what has been done for string field theory, but using now ingredients that arise from the geometry of a supermanifold, we can construct 
 multilinear products of forms. Some of them have precisely the same form of those coming from superstring field theory - 
 constructed in terms of wedge products and PCO's insertions -, but on the other hand the richness that emerges from the geometry of forms on supermanifolds 
 leads to new products, turning the exterior algebra of forms into a non-associative algebra 
 generalizing the above-mentioned $A_\infty$-algebra construction \cite{Erler}. This completes the construction of new products 
 of forms endowed with new algebraic properties. Applications of these new structures are still premature, 
 but can be foreseen in several directions. 
  
 The paper is organized as follows: in sec. 2, we review the basic ingredients in the theory of
 forms on supermanifolds. In sec. 3, we discuss the differential operators $d$ and $\eta$. In sec. 4, we 
 review a useful construction of the PCO $Z$ and 
 we show some computations as illustrative examples. In sec. 5, we introduce and discuss the $A_\infty$-algebra of forms on a supermanifold; in sec. 5.1 we compute the $M_3$ product in terms of $M_2$ products. In sec. 5.2 we give some explicit examples and finally in App. A we provide some useful computations. 
 
\section{Forms on Supermanifolds and their Local Representation}

In a supermanifold ${\mathcal {SM}}^{(n|m)}$, locally described by the coordinates $(x^a, \theta^\alpha)$, with $a=1,
\dots, n$ and $\alpha =1, \dots, m$,
we consider the spaces
of forms
$\Omega^{(p|r)}$ \cite{CDGM}. A given $(p|r)$-form $\omega$ can be expressed in terms of local generators as a formal sum as follows
\begin{eqnarray}\label{inNBA}
\omega &=& \sum_{l,h,r} \omega_{[a_1 \dots a_l] (\a_{1} \dots \a_{h}) [\b_{1} \dots \b_{r}]}(x,\theta) \times  \nonumber \\&\times& dx^{a_1} \dots dx^{a_l}
(d\theta^{\a_1})^{u(\a_1)} \dots (d\theta^{\a_h})^{u(\a_h)} \delta^{^{(g(\beta_1)})}(d\theta^{\b_1})   \dots \delta^{^{(g(\beta_r)})}(d\theta^{\b_r})
\end{eqnarray}
where $u(\alpha) \geq 0 $ is the power of the monomial $d\theta^\alpha$ and where $g(\alpha)$ denotes the differentiation degree of the Dirac delta form with respect to the $1$-form $d\theta^\alpha$.
Namely, $\delta^{(g(1))}(d\theta^1)$ is the $g(1)$-derivative of the Dirac delta with respect to the variable $d\theta^1$.\footnote{We recall that $d\theta^1 \delta^{^{(g(1))}}(d\theta^1) = - g(1) \delta^{^{(g(1)-1)}}(d\theta^1)$.}
The total form degree of $\omega^{(p|r)}$ is
\begin{equation}\label{inNBB}
l + \sum_{j=1}^h u(\alpha_j) - \sum_{k=1}^r g(\b_k) = p \in \mathbb{Z}\,, ~~~~~~\{\a_1, \dots, \a_r\} \neq \{\b_1, \dots, \b_r\} ~~~ \forall i=1,\dots,h.
\end{equation}
Note that each $\alpha_l$ in the above summation must be different from any $\beta_k$, otherwise the
degree of the differentiation of the Dirac delta function could be reduced and the corresponding $1$-form $d\theta^{\a_k}$ could be removed from the basis. The picture number $r$ corresponds to the number of Dirac delta forms.
 The components  $\omega_{[i_1 \dots i_l] (\a_{1} \dots \a_{m}) [\b_{1} \dots \b_{r}]}(x,\theta)$  of $\omega$ are superfields\footnote{Symmetrization and (anti)-symmetrization correspond to the parity of the generators involed.}, \emph{i.e.} local sections of the structure sheaf of the supermanifold $\mathcal{SM}.$

The graded wedge product is defined as usual
\begin{equation}\label{inNA}
\wedge: \Omega^{(p|r)} (\mathcal{SM}) \otimes \Omega^{(q|s)} (\mathcal{SM}) \longrightarrow
\Omega^{(p+q|r+s)} (\mathcal{SM})\,.
\end{equation}
where $0\leq p,q \leq n$ and $0\leq r,s \leq m$. Due to the anticommuting properties of the Dirac delta forms
$\delta(d\theta^\alpha) \delta(d\theta^\b) = - \delta(d\theta^\b) \delta(d\theta^\a)$
this product can be set equal to zero, if two delta forms has the same $d\theta$ as argument.\footnote{In addition,
it follows $\delta(d\theta^\a) \delta'(d\theta^\a) =0$, and consequently $\delta^{(p)}(d\theta^\a) \delta^{(q)}(d\theta^\a))  =0$
for any derivative $p,q$ of the Dirac delta forms.}

Actually, supergeometry allows for an even richer scenario: introducing the \emph{inverse forms} as in \cite{Catenacci:2018xsv}, the complex made of the spaces $\Omega^{(p|q)}$
gets extended as follows
\begin{enumerate}
\item  For picture $q=0$, there are new superforms in $\Omega^{(p|0)}$ that can also carry a \emph{negative} form degree $p<0$. Locally, for a supermanifold of dimension $n|m$ we will have expressions of this kind
\begin{eqnarray}
\label{inNAA}
\omega^{(p|0)} = \sum_{l = 0}^n \sum_{r =0 }^m \sum_{a_i = 1 }^{n} \sum_{\alpha_j =1}^m \omega_{[a_1 \dots a_l] (\alpha_1 \dots \alpha_r)}(x,\theta) { dx^{a_1} \dots dx^{a_l}}{(d\theta^{\alpha_1})^{u(\alpha_1)} \dots (d\theta^{\alpha_r})^{u(\alpha_r)} }
\end{eqnarray}
together with the constrain $p = l + \sum_{r} u (\alpha_r)$ and where $u (\alpha_j) \in \mathbb{Z}$ is the power of the monomial $d\theta^{\alpha_r}$, that can now take also negative values. For example, on $\mathbb{C}^{{1|1}}$, one might consider forms of degree $-1$, having the following form 
\begin{equation}
\omega^{(-1|0)}_{\mathbb{C}^{1|1}} = \omega_0 (x,\theta) \frac{1}{d\theta} + \omega_1 (x,\theta) \frac{dx}{d\theta^2} 
\end{equation}
\item Notice that, in general, whenever the supermanifold has fermionic dimension greater that $1$, each space $\Omega^{(p|0)}$ for $p \in \mathbb{Z}$ has an \emph{infinite} number of generators - even for $p \geq 0$. Consider for example the case of $\mathbb{C}^{1|2}:$ allowing for inverse forms, beside $1$, the space of $\Omega^{(0|0)}_{\mathbb{C}^{1|2}}$ is generated by all of the expressions of the kind $d\theta_1^{p_1} d\theta_2^{p_2}$ with $p_1 = - p_2$ and $dx d \theta_1^{p_1} d\theta_2^{p_2}$ with $p_1 + 1 =  - p_2$ where  $ p_1 , p_2 \in \mathbb{Z}$.  
\end{enumerate}

\noindent The spaces $\Omega^{(p|r)}$ with intermediate picture, namely when $0< r < m$,
are infinitely generated:
\begin{enumerate}
\item There might appear derivatives of Dirac delta forms of any order 
$\delta^{g(\alpha_l)}(d\theta^{\alpha_l})$ which reduce the form degree.
\item There might be any powers of $(d\theta^\alpha)^{u(\alpha)}$ different from those contained into the Dirac delta's, namely $\prod_{r=0}^l (d\theta^{\alpha_r})^{u(\alpha_r)} \prod_{s=l+1}^m \delta^{g(s)}
(d\theta^{\alpha_s})$ where $\alpha_i \neq \alpha_j$ with $i=1, \dots, l$ and $j=l+1, \dots, m$. 
The powers $u(\alpha)$ can be positive or negative, while the $g(s)$ are non-negative. 
\end{enumerate}

Finally, the complex $\Omega^{(p|m)}$ is bounded from above, since there are no other form above the top form $\Omega^{(n|m)}$ and, at a given form degree, each space $\Omega^{(p|m)}$ is finitely-generated.
\\

The \emph{odd} differential operator $d$ maps forms of the type $\Omega^{(p|r)}$ into
forms of the type $\Omega^{(p+1|r)}$ increasing the form number without changing the picture. The action of the differential operator $d$ on the Dirac delta functions is by chain rule, namely $\delta (f (d\theta))) = \delta^\prime( f(d\theta)) d f(d\theta)$, so that, in particular, $d\delta (d\theta ) = 0.$ \\

We now focus on a supermanifold of dimension $(1|2)$ for simplicity and we consider the form spaces
$\Omega^{(p|q)}$ with $0 \leq q \leq 2$. A simple but non-trivial example of supermanifold of dimension $(1|2)$ is the projective superspace $\mathbb{P}^{1|2}$ over the complex numbers. For a detailed discussion about the geometry of projective superspaces see for example \cite{CN}.
It is defined starting with two patches $U_0$ and $U_1$ and the mapping of the coordinates $z_0, \theta^\a_0$
to the coordinates $z_1, \theta^\a_1$ is given by the homolorphic transition functions
\begin{eqnarray}
\label{transA}
z_0 \longmapsto z_1 =\frac{1}{z_0}\,, ~~~~\theta^\a_0 \longmapsto \theta^\alpha_1 = \frac{\theta^\a_0}{z_0} \quad \alpha = 1, 2.
\end{eqnarray}
Its Berezinian bundle is generated by the section $\omega^{(1|2)} = dz \delta(d\theta^1) \delta(d\theta^2)$
which is globally defined, and indeed the supermanifold $\mathbb{P}^{1|2}$ is an example of Calabi-Yau supermanifold \cite{1DCY}. For $q=0$ and $p \in \mathbb Z$, we call the
space $\Omega^{(p|0)}_{\mathbb{P}^{1|2}}$, the space of {\it superforms} . For
$q=1$ and $p \in \mathbb Z$, we call  $\Omega^{(p|1)}_{\mathbb{P}^{1|2}}$
the space of {\it pseudoforms} and finally,
for $q=2$ and $p \leq 1$, we call  $\Omega^{(p|2)}_{\mathbb{P}^{1|2}}$
the space of {\it integral} forms.

\section{The Differential Operators $d$ and $\eta$}

We now work over the supermanifold $\proj {1|2}$. There are two differential operators acting on the complex of forms: the obvious one is the usual
odd differential $d$
\begin{eqnarray}
\label{diff_A}
d: \Omega^{(p|q)}_{\proj {1|2}} \longrightarrow \Omega^{(p+1|q)}_{\proj {1|2}}
\end{eqnarray}
As already stressed, it increases the form number, but it does not change the picture. We now introduce another differential operator that will be used in what follow, but first we need some auxiliary material. 

Let $D$ be a vector field in the tangent bundle of the supermanifold $\mathcal{T}_{\proj {1|2}}$. In local coordinates is expressed as 
\begin{eqnarray}
\label{ant_AA}
D = D^z(z,\theta)  \frac{\partial}{\partial z} + D^\a(z,\theta) \frac{\partial}{\partial \theta^\a}. 
\end{eqnarray}
Then, for a \emph{constant odd vector field}
one has 
\begin{equation}
D = D^1 \frac{\partial}{\partial \theta^1} + D^2 \frac{\partial}{\partial \theta^2}
\end{equation}
with $D^1, D^2 \in \mathbb{C}.$ Clearly, two odd vector fields $D$ and $D'$,
are linearly independent if $\det(D,D') = D^1 D^{2'} - D^{1'} D^2 \neq 0$.

\noindent In general, given a vector field $D$, one can define the inner product $\iota_D$ which acts as
\bear
\label{cartan_AA} \xymatrix@R=1.5pt{
\iota_D : \Omega^{(p|q)}_{\proj {1|2}} \ar[r] & \Omega_{\proj {1|2}}^{(p-1|q)}\\
\omega \ar@{|->}[r] & \iota_D (\omega )},
\eear
where $\iota_D (\omega) (X_1, \ldots, X_{p-1}) \defeq \omega (D, X_1, \ldots, X_{p-1})$. For $D \defeq D^\alpha \partial_{\theta^\alpha},$ an odd constant vector field as requested above one has, in particular that
\bear
\iota_{D} (d\theta^\alpha) = d\theta^\alpha (D) = D^\alpha.
\eear
Also, notice that it satisfies the usual Cartan algebra
\begin{eqnarray}
\label{cartan_A}
{\cal L}_D = [d, \iota_D]\,,  ~~~~~~
[{\cal L}_D, \iota_{D'}] = \iota_{\{D,D'\}}\,, ~~~~~
\{\iota_D, \iota_{D'}\} =0\,,
\end{eqnarray}
where $\mathcal{L}_{D}$ is the Lie derivative along $D$. We stress that in the first identity the commutator $[\cdot, \cdot]$ replaces the anticommutator $\{\cdot, \cdot\}$ since
the differential operator $\iota_D$ has parity opposed to that of $D$ - so that if $D$ is odd, one has $|\iota_D| = 0$ - and the differential $d$ is odd.
For constant $D$ and $D'$, $\{D,D'\}=0$ and $\iota_D^2 \neq 0$.

As we learnt from string theory (see \cite{Berkovits:1994vy}) there is another interesting odd differential operator which can be defined from $\iota_D$ (again for $D$ an odd constant vector field) upon using the Euler representation of the sine: \footnote{We use the normalization such that $\delta(x) = \int_{-\infty}^{\infty} e^{i t x} dt$ and $\Theta(x) = - i  \lim_{\epsilon \rightarrow 0} \int_{-\infty}^{\infty} \frac{e^{i t x}}{t + i \epsilon} dt$, and $\Theta'(x) = \delta(x)$. In this way, in order to match the correct assignments we need the factor $-2$ in the definition of $\eta$ in (\ref{diff_B})
. } 
\begin{eqnarray}
\label{diff_B}
\eta = - 2 \Pi \lim_{\epsilon \rightarrow 0} \sin(i \epsilon \iota_D): \Omega_{\proj {1|2}}^{(p|q)} \rightarrow \Omega_{\proj {1|2}}^{(p+1|q+1)}
\end{eqnarray} where $\Pi$ is the parity-change functor (see \cite{Catenacci:2018xsv}) that simply changes the parity of the 
expression to which is applied, without affecting any other property. Acting with $\eta$ on the inverse forms $1/d\theta^\alpha$, we 
have 
\begin{eqnarray}
\label{diff_BA}
\eta \left(\frac{1}{d\theta^\a} \right) = \delta(d\theta^\a)\,, 
\end{eqnarray}
where $1/d\theta^\alpha$ is even and $\delta(d\theta^\a)$ is odd according to the axioms defining these 
distributions.

\noindent The differential operator $\eta$ acts as follows
\begin{eqnarray}
\label{cartan_B}
\left \{
\begin{array}{lrr}
  \eta \Big( \frac{1}{(d\theta^\alpha)^p} \Big) =
\frac{(-1)^{p-1}}{(p-1)!} \delta^{(p-1)}(d\theta^\a) & & p>1,
   \\ ~~\\
   \eta \Big( (d\theta^\alpha)^p \Big) =0 & & p\geq 0,
   \\ ~~ \\
 \eta \Big( \delta^{(p)}(d\theta^\alpha) \Big) =0 & & p \geq 0,
\end{array}
\right.
\end{eqnarray}
whilst it does not act on the differentials of even coordinates. It is easy to verify that
\begin{eqnarray}
\label{cartan_C}
\eta^2 = 0\,, ~~~~~~ \{d, \eta\} =0, 
\end{eqnarray}
as in string theory \cite{Berkovits:1994vy}. In addition, $\eta$ is a graded-derivation with respect to the exterior algebra
\begin{eqnarray}
\label{cartan_BA}
\eta(\omega_A \wedge \omega_B) = \eta(\omega_A) \wedge \omega_B + (-1)^{|\omega_A|} \omega_A \wedge \eta(\omega_B)\,. 
\end{eqnarray}
where $\omega_A$ and $\omega_B$ are forms of the complex $\Omega^{p|q}$. 

The operator $\eta$ has been introduced in string theory to select the Small Hibert Space (SHS) inside the Large Hilbert space (LHS). 
As discussed in \cite{Catenacci:2018xsv}, the LHS 
for a supermanifold is constructed by adding the inverse forms (which are still distribution-like forms) to $\Omega^{p|q}$ with $q<m$. 
Thus, the equation
\begin{eqnarray}
\label{diff_BA}
\eta (\omega^{(p|q)})=0
\end{eqnarray}
selects the forms that are in the SHS.

\section{The PCO $Z_D$}

When computing amplitudes in string theory, one introduces the {\it picture changing operators} (PCO) in order to change the
picture of the vertex operators as to saturate the superghost charges according to the anomaly cancellation.
In string theory the PCO are independent of the position of their insertion into the amplitude since
the string fields are on-shell. On the other hand, in the string field theory action, when the string fields are off-shell, the position of the PCO really matters. The consequence of a ``wrong'' choice is the loss of the gauge invariance of the theory.\\
In the present supergeometric framework, operators analogous to the PCO of string theory can be defined as acting on the complex of forms $\Omega^{(p|q)}$, moving from one picture to another and leaving the form number unchanged. In addition, it can be shown that they are isomorphisms in de Rham cohomology 
\cite{CDGM}.  

With these preliminary remarks, we can define the PCO $Z_D$ (see for example \cite{CCGhd}) and ancillary operators as follows
\begin{eqnarray}
\label{PCO_A}
&&Z_D \defeq \{d, - i \Theta(\iota_D)\}\,, 
~~~~  \Theta(\iota_D) \defeq - i \lim_{\epsilon \rightarrow 0}\int^{\infty}_{-\infty} dt \frac{e^{i t \iota_D}}{t + i \epsilon}\,, 
~~~~  \delta(\iota_D) \defeq \int^{\infty}_{-\infty} dt e^{i t \iota_D}. 
\end{eqnarray}
The latter two act as follows on a certain polynomial functions $f$ of the $d\theta$'s:
\begin{align}
& \Theta(\iota_D)  f(d\theta^\a) = - i \lim_{\epsilon \rightarrow 0}\int^{\infty}_{-\infty} dt 
 \frac{f(d\theta^\a + i t D^\a)}{ t + i \epsilon}, \nonumber \\
 & \delta(\iota_D)  f(d\theta^\a) =  \int^{\infty}_{-\infty} dt 
 {f(d\theta^\a + i t D^\a)}. 
\end{align}
We add the subscript $D$ to the PCO to recall that it depends on the choice of the
odd vector $D$. Note that formally the operator $Z_D$ is {exact}, \emph{i.e.} $Z_D = \{ d, - i \Theta (\iota_D) \}$, but $\Theta(\iota_D)$ acting on $\Omega^{(p|q)}$ 
brings from the SHS to the LHS. However, by computing the variation of $Z_D$ by a change of $D$ (that is 
$D \rightarrow D + \delta D$), $Z_D$ transform ad $\delta Z_D = \{d, \delta(\iota_D) \delta D^\a \iota_\a \}$ 
which is exact and $\delta(\iota_D) \delta D^\a \iota_\a$ acts on $\Omega^{(p|q)}$ staying into the SHS.

The two operators $\Theta(\iota_D)$ and $\delta(\iota_D)$ act as follows
\begin{eqnarray}
\label{ant_A0}
\Theta(\iota_D): \Omega^{(p|q)} \longrightarrow \Omega^{(p-1|q-1)}\,,
~~~~~~~~~~~
\delta(\iota_D): \Omega^{(p|q)} \longrightarrow \Omega^{(p|q-1)}\,.
\end{eqnarray}
Both reduce the picture, but the first one reduces also the form degree. This is consistent
with the usual relation between the Heaviside $\Theta$ function and the Dirac delta $\delta$. In the
case of $q=0$, acting on zero-picture forms, they vanish. Notice also that, differently from $\eta$, they are \emph{not}
derivations of the exterior algebra of forms. In app. A, we will give some explicit computations in order to 
clarify the action of $\Theta(\iota_D)$ and $\delta(\iota_D)$ operators. 

Note that if $m>1$, we have $m$ linear independent odd vectors $D_i$, therefore we can define a PCO
corresponding to each of them. In addition, we have a map between integral forms and superforms as
the product of all independent PCO's
\begin{eqnarray}
\label{PCO_B}
Z_{max} = \prod_{i=1}^m \Big\{d, - i \Theta(\iota_{D_i})\Big\}: \Omega^{(p|m)} \longrightarrow \Omega^{p|0}\,.
\end{eqnarray}
Due to the anticommutative properties of $\Theta(\iota_{D_i})$, it is easy to prove that $Z_{max}$ does not depend on the choice of the odd
vector fields $D_i$.


In general, for $\omega \in \Omega^{(p|q)}$ when $q=0,1$, the absence of inverse forms is guaranteed if
\begin{eqnarray}
\label{ant_DA}
\eta(\omega) =0\, \qquad \omega \in \Omega^{(p|q)}, \;  q=0,1.
\end{eqnarray}
For $q=2$, $\eta$ acts trivially since there is no room for inverse forms. For $q=0,1$, we observe
\begin{eqnarray}
\label{ant_DB}
\eta( Z_D ( \omega ) )  &=& \eta \left( \{d, -i \Theta(\iota_D) \} (\omega) \right)
=\eta \Big( -i d \Theta(\iota_D) \omega -i \Theta(\iota_D) d \omega \Big) \nonumber \\
&=&- i( - d \eta \Theta(\iota_D) \omega + \{\eta, \Theta(\iota_D)\} d\omega - \Theta(\iota_D) \eta (d\omega) )\nonumber \\
&=&  - i (  - d \{\eta, \Theta(\iota_D) \}\omega + d \Theta(\iota_D) \eta(\omega)
+ \{\eta, \Theta(\iota_D)\} d\omega - \Theta(\iota_D) \eta (d\omega) ) \nonumber \\
&=& Z_D ( \eta(\omega)) =0\,.
\end{eqnarray}
where we have used $\{\eta, - i \Theta(\iota_D)\} = 1$ for any $D^\a$ and $\{d, \eta\}=0$. Therefore,
the PCO-transformed $\omega$, namely $Z_D (\omega )$, 
is independent of inverse form if $\omega$ is independent.


\section{$A_{\infty}$-algebra of forms}

In the complex of forms there is a natural bilinear map represented by the usual exterior product
\begin{eqnarray}
\label{wedA}
\wedge:  
\Omega^{(p|q)}\otimes \Omega^{(p^{\prime}|q^{\prime})}\rightarrow\Omega^{(p+p^{\prime}|q+q^{\prime})}\,. 
\end{eqnarray}
Note that, in general, the wedge product changes the form degree and the picture according to the previous formula.
In addition, working over $\proj {1|2}$, for $q=1, q'=2$ or $q=2, q'=1$ or $q=q'=2$, the products 
are trivial. The differential $d, \eta$ are derivations of the exterior algebra. Also, the exterior product 
can be extended to consider the inverse forms letting $p <0$ for $q=0$. 

We will show how to construct new bilinear maps that change the picture according to a different prescription.  
For example, in string field theory, the role of a $(1|1)$ form is played by a 
string vertex operator with ghost number 1 and picture number 1. The authors in \cite{Erler} constructed a new bilinear map 
which take the product of these two vertices into a new vertex with quantun numbers 
$(2|1)$, namely ghost number two and the same picture. This new product is not associative and it leads to a structure of $A_\infty$-algebra: we will show how this structure arises geometrically from superforms defined on a supermanifold. 

We briefly recall the definition of $A_\infty$-algebra (see for example \cite{Aspinwall, Keller} for further details) and then we give an explicit realization in terms 
of the complex of forms. \\
An $A_{\infty}$-algebra \cite{Keller} is a graded vector space $A=\bigoplus_{p\in \mathbb{Z}} A^{p}$  with $p\in\mathbb{Z},$
endowed with graded maps (homogeneous and linear) 
\begin{eqnarray}
\label{wedB}
M_{n}: 
A^{\otimes n}\rightarrow A, ~~~~ n\geq1
\end{eqnarray}
of degree $2-n$ satisfying the relations:

\begin{enumerate}
\item $M_{1}M_{1}=0$, \emph{i.e.} the pair $(A, M_1)$ is a differential complex;

\item $M_{1}M_{2}=M_{2}\left(  M_{1}\otimes1+1\otimes M_{1}\right)$, \emph{i.e.} $M_1$ is derivation with respect to the multiplication $M_2$.

\item $M_{2}\left(  1\otimes M_{2}-M_{2}\otimes1\right)  =M_{1}M_{3}%
+M_{3}\left(  M_{1}\otimes1\otimes1+1\otimes M_{1}\otimes1+1\otimes1\otimes
M_{1}\right)  $

\item for $n\geq1$ we have $\sum(-1)^{r+st}M_{u}(1^{\otimes r}\otimes
M_{s}\otimes1^{\otimes s}) =0 $. The sum is over the decompositions $n=r+s+t$ and
$u=1+r+t.$
\end{enumerate}
Note that $\left(  M_{1}\otimes1+1\otimes M_{1}\right)  $ acts as the usual odd
differential $d$ on the tensor algebra:%
\[
\left(  M_{1}\otimes1+1\otimes M_{1}\right)  (x\otimes y)=M_{1}(x)\otimes
y+(-1)^{|x|}x\otimes M_{1}(y)
\]
In the present case, we consider a double-graded vector space
\begin{eqnarray}
\label{wedC}
A = \bigoplus_{p,q} A^{(p|q)} = \bigoplus_q A^{(\bullet|q)}
\end{eqnarray}
where the first number denote the form degree and the second the picture number. 
In our case, we take $A^{(\bullet|q)}$ to be $\Omega^{(\bullet|q)}$, the complex of forms at fixed picture number $q$ (note that, strictly speaking this is a complex of sheaves of vector spaces). 
The form number is an integer for $q < m$, where $m$ is the maximum number of fermion dimensions. 
For $q =m$, the form number is bounded by $n$, the maximum number of bosonic dimension. 
We define the graded maps 
$
M^{(-l)}_{n}: 
A^{\otimes n}\rightarrow A$, with  $n\geq1$
as follows 
\begin{eqnarray}
\label{wedBA}
M^{(-l)}_{n}: \Omega^{p_1|q_1} \otimes \dots \otimes  \Omega^{p_n|q_n} \longrightarrow 
\Omega^{p'| q'}\,, ~~~~~ p' =  \sum_{i=1}^n p_i, ~~~~ q' =  \sum_{i=1}^n q_i - l\,. 
\end{eqnarray}
We notice that the map  $M^{(-l)}_{n}$ is defined as to lower the {sum of the} pictures by $l$, \emph{i.e.} if $\omega_1, \ldots, \omega_n$ have picture $q_1, \ldots, q_n$ respectively, the form $M^{(-l)}_n (\omega_1 \otimes \ldots \otimes \omega_n)$ is of picture $q' = \sum_{i=1 }^n q_i  - l$.  

For $n=1$ there are the two 
representatives, $d$ and $\eta$. Both are graded linear maps:
\begin{eqnarray}
\label{wedD}
M_1^{(0)} \defeq d\,, ~~~~~~~~
M_1^{(1)} \defeq \eta\,. 
\end{eqnarray}

In the simple case of two fermionic dimensions, namely for example working in $\mathbb{P}^{1|2}$ or 
also $\mathbb{A}^{0|2}$ (where $q\leq 2$), we define the $2$-products as
\begin{eqnarray}
\label{wedBA}
M_{2}^{\left(  0\right)  } & \defeq&  \wedge \,,\nonumber\label{prod_A}\\
M_{2}^{\left(  -1\right) } & \defeq& \frac{1}{3}\left[
Z_{D} M_2^{(0)} + M_2^{(0)} (Z_D \otimes 1 + 1 \otimes Z_D) \right] 
\nonumber\\
M_{2}^{\prime \left(  -1\right) } & \defeq& \frac{1}{3}\left[
Z_{D^\prime} M_2^{(0)} + M_2^{(0)} (Z_{D^\prime} \otimes 1 + 1 \otimes Z_{D^\prime}) \right] 
\nonumber
\end{eqnarray}
\begin{eqnarray}
M_{2}^{\left(  -2\right)  } & \defeq &
\frac{1}{3^{2}}\left[
Z_{D^{\prime}} M_2^{(-1)} + M_2^{(-1)}(Z_{D^{\prime}} \otimes 1 + 1 \otimes Z_{D^{\prime}}) \right]
\nonumber \\
&= &
\frac{1}{3^2}
\left[ Z_{D^{\prime}} Z_D M_2^{(0)} + 
Z_{D^{\prime}} M_2^{(0)} (Z_{D} \otimes 1 + 1 \otimes Z_{D}) 
+ 
Z_{D} M_2^{(0)} (Z_{D^\prime} \otimes 1 + 1 \otimes Z_{D^\prime}) \right. \nonumber \\
&&\left. ~~~~~~~ +  
M_2^{(0)}
\left(Z_{D^{\prime}} Z_D  \otimes 1 + Z_{D^{\prime}} \otimes Z_D + Z_{D} \otimes Z_{D^{\prime} }
+ 1 \otimes Z_{D^{\prime}} Z_D  \right) 
\right]
\end{eqnarray}
They are built starting from the (graded supersymmetric) wedge product $M_{2}^{\left(  0\right)  }=  \wedge$ and by 
inserting the PCOs into the product in a symmetric way respecting the tensor structure.  Note that there are two equivalent $2$-products reducing the picture by one, \emph{i.e.} $M_2^{(-1)}$ and $M^{\prime (-1)}_2$. This is due to 
the presence of two independent fermionic directions. 

Notice that in the case $q=q^{\prime}=1$ the product  $M_{2}^{\left(
-1\right)  }$ maps $\Omega^{(p|1)}\otimes\Omega^{(p^{\prime}|1)}\rightarrow
\Omega^{(p+p^{\prime}|1)}$ and in the case $q=q^{\prime}=2,$ the product
$M_{2}^{\left(  -2\right)  }$ maps $\Omega^{(p|2)}\otimes\Omega^{(p^{\prime}%
|2)}\rightarrow\Omega^{(p+p^{\prime}|2)}.$ These products preserve the picture
and are bilinear maps from $\Omega^{(\bullet|q)}\otimes\Omega^{(\bullet|q)}\ $ to
$\Omega^{(\bullet|q)}$. 

\newcommand{\MM}{\mathbf M}

It is convenient to rewrite these maps using the notion of \emph{coderivations}, in terms of which we can use the graded commutators, denoted here and in the following 
as $[\,\cdot \, ,\, \cdot \,]$, simplifying the computations (see \cite{Erler} for details). Given a multilinear map 
$\Delta_n : A^{\otimes n} \rightarrow A$ of the graded vector space $A$, we define the \emph{associated coderivation} ${\mathbf \Delta}_n : A^{\otimes N} \rightarrow A^{\otimes (N-n+1)} $ for any $N\geq n$ as follows:\footnote{In the following the coderivation associated to a map will be noted by same character, but in boldface style.} 
\bear
\label{wedD}
{\mathbf \Delta}_n \defeq \sum_{k=0}^{N-n} 1^{\otimes (N-k-n)} \otimes \Delta_n \otimes 1^{\otimes k}
\eear
acting on the spaces $A^{\otimes N \geq n}$.  Note that if $N=n$, one has that ${\mathbf \Delta}_n = \Delta_n$, so in this particular context one can see that a coderivation ``extends'' a certain multilinear map $\Delta_n$ to higher tensor powers of $A$ by suitably tensoring it with the identity map. \\
It can be seen that the commutator of two coderivations associated to the multilinear maps $\Delta_m : A^{\otimes m} \rightarrow A$ and $\Delta_n^\prime : A^{\otimes n} \rightarrow A$ respectively, is 
the coderivation associated with the commutator of the maps $\Delta_m$ and $\Delta^\prime_n$. The commutator acts as
\bear
[\Delta_m, \Delta^\prime_n] : A^{\otimes  (n+m -1)} \longrightarrow A
\eear
and it is explicitly given by 
\begin{eqnarray}
\label{wedE}
&&[\Delta_m, \Delta^\prime_n] =\\
&& =\Delta_m \left [ \sum_{k=0}^{m-1} 
1^{\otimes (m-k-1)} \otimes \Delta^{\prime}_n \otimes 1^{\otimes k} \right ] - 
(-1)^{|\Delta_m||\Delta^\prime_n|} \Delta^{\prime}_n  \left [
 \sum_{k=0}^{n-1} 
1^{\otimes (n-k-1)} \otimes \Delta_m \otimes 1^{\otimes k} \right ].  \nonumber 
\end{eqnarray}
Notice that this is well-defined. Indeed, looking for example at the first bit of the commutator in the previous equation, one has that $\sum_{k=0}^{m-1} 
1^{\otimes (m-k-1)} \otimes \Delta^{\prime}_n \otimes 1^{\otimes k} : A^{\otimes (m+n-1)} \rightarrow A^{\otimes m}$ and hence it maps tensors in the domain of the multilinear map $\Delta_m$, as it should. Similar story goes on for the second bit.\\
 This notation is an economical way to keep track of the $A_\infty$-relations. 
For example, the first {three} defining relations of the $A_{\infty}$-algebra are re-written as follows
\begin{eqnarray}
\label{wedF}
[{\mathbf M}_1, {\mathbf M}_1] =0\,, ~~~~~
[{\mathbf M}_1, {\mathbf M}_2] =0\,, ~~~~~
\frac12 [{\mathbf M}_2, {\mathbf M}_2] + [{\mathbf M}_1, {\mathbf M}_3]  = 0\,. ~~~~~
 \end{eqnarray}
If we denote by ${\mathbf M}_2^{(0)}$ the coderivation constructed in terms of $M_2^{(0)}$, \emph{i.e.} we consider the case $l=0$,
we have an associative DG-algebra, where ${\mathbf M}^{(0)}_1$ is the differential $d$.   
We have the relations 
\begin{eqnarray}
\label{wedF}
[{\mathbf M}^{(0)}_1, {\mathbf M}^{(0)}_1] =0\,, ~~~~~
[{\mathbf M}^{(0)}_1, {\mathbf M}^{(0)}_2] =0\,, ~~~~~
\frac12 [{\mathbf M}^{(0)}_2, {\mathbf M}^{(0)}_2]  = 0\,, ~~~~~
 \end{eqnarray}
with ${\mathbf M}^{(0)}_n =0$ for $n>2$.\\ 
Let us consider now the case $l=1$. Using the commutators, we can rewrite the 
second coderivation $\MM^{(-1)}_2$ of (\ref{wedBA}) as follows
\begin{eqnarray}
\label{wedG}
{\mathbf M}^{(-1)}_2 = [Z_D, {\mathbf M}^{(0)}_2]\,.    
\end{eqnarray}
It is easy to verify the first $A_\infty$-relation 
\begin{eqnarray}
\label{wedH}
[\MM^{(0)}_1, {\mathbf M}^{(-1)}_2] = 
[\MM^{(0)}_1,[Z_D, {\mathbf M}^{(0)}_2]] = 
- [Z_D,[ {\mathbf M}^{(0)}_2,\MM^{(0)}_1]] - [\MM^{(-1)}_2,[{\mathbf M}^{(0)}_1, Z_D]] =0
\end{eqnarray}
since $[ {\mathbf M}^{(0)}_2,\MM^{(0)}_1] =0$ by (\ref{wedF}) and 
$[{\mathbf M}^{(0)}_1, Z_D]=0$ since the PCO is $d$-closed. On the other 
hand, by a simple computation, one can see that the associativity of $\MM_2^{(-1)}$ is violated 
\begin{eqnarray}
\label{wedI}
[{\MM}^{(-1)}_2, {\MM}^{(-1)}_2] =  {\mathbf \Delta}^{(-2)}_3 \neq 0\,, 
\end{eqnarray}
where the violation comes from a $3$-product $\Delta^{(-2)}_3$. Using 
again the relations (\ref{wedF}) and (\ref{wedG}), we have 
\begin{eqnarray}
\label{wedK}
[\MM_1^{(0)}, { \mathbf \Delta}^{(-2)}_3 ] =  
[\MM_1^{(0)}, [{\MM}^{(-1)}_2, {\MM}^{(-1)}_2]] = - 2 [\MM_2^{(-1)}, [{\MM}^{(0)}_1, {\MM}^{(-1)}_2]] =0 
\end{eqnarray}
 which implies that ${ \mathbf \Delta}^{(-2)}_3$ is $d$-closed. If ${ \mathbf \Delta}^{(-2)}_3$ 
 were formally exact, namely it exists an $\MM_3^{(-2)}$ such that 
 \begin{eqnarray}
\label{wedKA}
{ \mathbf \Delta}^{(-2)}_3 = [\MM_1^{(0)}, \MM_3^{(-2)}]\,, 
\end{eqnarray}
 then it would follow 
\begin{eqnarray}
\label{wedL}
\frac12 [{\MM}^{(-1)}_2, {\MM}^{(-1)}_2] + [\MM_1^{(0)},\MM^{(-2)}_3] =0,   
\end{eqnarray}
which is the starting point to build the $A_\infty$-algebra. The proof that ${ \mathbf \Delta}^{(-2)}_3$ is 
indeed exact is deferred to the next subsection where $\Theta(\iota_D)$ is used to compute $\MM^{(-2)}_3$. 
Repeating this analysis, one finds a series of coderivations $\MM^{(1-n)}_n$ satisfying the 
$A_\infty$-relations. \\

As stated at the beginning of the section, there is another differential operator $M_1^{(1)} =\eta$, acting 
on the space of forms as a graded-derivation of the exterior algebra, namely with respect to $M_2^{(0)}$. Getting back to the coderivation notation, this can be 
expressed as 
\begin{eqnarray}
\label{assoA}
[\MM_1^{(1)}, \MM_2^{(0)}] =0\,. 
\end{eqnarray}
We can check that $\MM_1^{(1)}$ is also a graded-derivation of $\MM_2^{(-1)}$ by 
observing that 
\begin{eqnarray}
\label{assoB}
[\MM_1^{(1)}, \MM_2^{(-1)}] = [\MM_1^{(1)}, [Z_D, {\mathbf M}^{(0)}_2]] = 
- [Z_D, [ {\mathbf M}^{(0)}_2, \MM_1^{(1)}]] 
- [{\mathbf M}^{(0)}_2, [\MM_1^{(1)}, Z_D]] = 0.
\end{eqnarray}
The right hand side vanishes since $\MM_1^{(1)}$ is a derivation of the wedge product by (\ref{assoA}), 
and the PCO commutes with $\eta$ as proven in (\ref{ant_DB}). This implies that the result of the 
product $\MM_2^{(-1)}$ is still in the SHS if both the forms on which it acts are in the SHS. In the same way, using the complete set of $A_\infty$-relations, one can prove that all coderivations $\MM_n^{(1-n)}$ commute with 
$M_1^{(1)} =\eta$. 

Finally, note that the set of products $\MM_n^{(1-n)}$ acts as 
\begin{eqnarray}
\label{assoC}
M_n^{(1-n)}: (\Omega^{(\bullet|1)})^{\otimes n} \longrightarrow \Omega^{(\bullet|1)}
\end{eqnarray}
since the wedge product of $n$ forms of $\Omega^{(\bullet|1)}$ has picture $n$, and the 
product $\MM_n^{(1-n)}$ lower the picture to $1$. 

Let us now consider the last product $M_2^{(-2)}$ in eq. (\ref{wedBA}). As discussed in the beginning, there are two 
possible products $M_2^{(-1)}$ and $M_2^{\prime (-1)}$ which can be constructed. They are associated to the two independent odd vectors $D$ and 
$D'$ that determine the two independent directions in the space of $d\theta$'s ({dually}). However, 
in picture 2, we find that there is only one possible product. Using again the coderivation notation we have
\begin{eqnarray}
\label{assoD}
\MM_2^{(-2)} = [Z_{D^\prime}, \MM^{(-1)}_2] = [Z_{D^\prime}, [Z_{D}, \MM^{(0)}_2]]
\end{eqnarray}
that satisfies  
\begin{eqnarray}
\label{assoE}
[\MM_1^{(0)}, \MM_2^{(-2)}] = 0\,, 
\end{eqnarray}
having made used of the Jacobi identity for the commutators and the formula (\ref{wedH}). A simple computation 
shows that the associativity of $\MM_2^{(-2)}$ is violated by a coderivation $\MM_3^{(-4)}$ as follows 
\begin{eqnarray}
\label{assoF}
\frac12 [{\MM}^{(-2)}_2, {\MM}^{(-2)}_2] + [\MM_1^{(0)},\MM^{(-4)}_3] =0\,.   
\end{eqnarray}
Again, as above, one can compute the $A_\infty$-relations between these products and 
the commutation relation with $\MM^{(1)}_1$. The final result gives the coderivations (and the 
corresponding multiproducts) $\MM_n^{(2 - 2n)}$. 
Note that these multiproducts act as follows 
\begin{eqnarray}
\label{assoG}
M_n^{(2 - 2n)}: (\Omega^{\bullet|2})^{\otimes n} \rightarrow \Omega^{\bullet|2}\,. 
\end{eqnarray}

We can generalize the set of products to multiplications of forms with \emph{different} pictures.  Indeed  
$\MM_2^{(-l)}$ acts on any type of forms regardless their picture, mapping them into 
a different complex, according to
\begin{eqnarray}
\label{assoH}
M_2^{(-l)}: \Omega^{(\bullet|p)} \otimes \Omega^{(\bullet|p')} \rightarrow \Omega^{(\bullet|p+p' -l)}.
\end{eqnarray}
This leads to study the commutator 
\begin{eqnarray}
\label{assoK}
[\MM_2^{(-l)}, \MM_2^{(-h)}] = \Delta^{-(l+h)}_{3}.
\end{eqnarray}
Since any $\MM^{(-l)}_2$ satisfies $[\MM_1^{(0)}, \MM^{(-l)}_2]=0$ (\emph{i.e.} the first $A_\infty$-relation), 
we have that $\Delta^{(-(l+h))}_{3} = [\MM_1^{(0)}, \MM^{(-(l+h))}_3]$ defining the new co-derivation 
$\MM^{(-(l+h))}_3$. It follows that the multiplicative structure can be extended to $\Omega^{(\bullet|\bullet)}$. 

\subsection{The Construction of $M_3^{(2 - 2h)}$}  

In this subsection we construct explicitly the $3$-product $M_3^{(2 - 2h)}$ for 
$h\geq 2$, following the suggestions in \cite{Erler}. 
We first review the construction of $M_3^{(- 2)}$ and then we derive the formula for 
the $3$-product which lowers the total picture changing by $4$.  

We first define the following coderivation 
\begin{eqnarray}
\label{cosA}
\widetilde {\MM}_{2}^{(-1)} = [ -i\, \Theta(\iota_D), \MM_2^{(0)} ]
\end{eqnarray}
which has the following properties 
\begin{eqnarray}
\label{cosB}
{\MM}_{2}^{(-1)} = [\MM^{(0)}_1, \widetilde {\MM}_{2}^{(-1)}]\,, 
~~~~~
 {\MM}_{2}^{(0)} = [\eta\, ,  \widetilde {\MM}_{2}^{(-1)}] \,, 
\end{eqnarray}
The  first equation means that ${\MM}_{2}^{(-1)}$ is formally exact. 
In the second equation $\eta$ is the differential operator defined in the previous section. 

Inserting the first equation into eq. (\ref{wedI}), we find that 
\begin{eqnarray}
\label{cosB}
{\mathbf \Delta}^{(-2)}_3 = \frac12 [ {\MM}_{2}^{(-1)}, [\MM^{(0)}_1, \widetilde {\MM}_{2}^{(-1)}]], 
\end{eqnarray}
which give us an explicit formula for the associator ${\mathbf \Delta}^{(-2)}_3$ in terms of the 
coderivations $\MM_2^{(-1)}$ and $\widetilde{\MM}_2^{(-1)}$. 
Using the Jacobi identity and the relation $[\MM_1^{(0)}, \MM_2^{(-1)}] =0$ in 
(\ref{wedF}), the right-hand side of (\ref{cosB}) can be re-written as 
\begin{eqnarray}
\label{cosBA}
{\mathbf \Delta}^{(-2)}_3 = -\frac12 [ {\MM}_{1}^{(0)}, [\MM^{(-1)}_{2}, \widetilde {\MM}_{2}^{(-1)}]]\,.
\end{eqnarray}
This concludes the proof that the associator ${\mathbf \Delta}^{(-2)}_3$ is formally $d$-exact. 
Finally,  using the definition in  (\ref{wedKA}) we have 
 \begin{eqnarray}
\label{cosC}
-\frac12 [ {\MM}_{1}^{(0)}, [\MM^{(-1)}_{2}, \widetilde {\MM}_{2}^{(-1)}]] + 
 [\MM^{(0)}_1, {\MM}_{3}^{(-2)}] =0\,.  
\end{eqnarray}
It follows that 
\begin{eqnarray}
\label{cosD}
\left [ {\MM}_{1}^{(0)}, \Big(\frac12  [\MM^{(-1)}_{2}, \widetilde {\MM}_{2}^{(-1)}  ]
+  {\MM}_{3}^{(-2)}\Big) \right ] =0,
\end{eqnarray}
from which we deduce 
\begin{eqnarray}
\label{cosE}
 {\MM}_{3}^{(-2)} = [\MM_1^{(0)}, \widehat \MM_3^{(-2)}] + \frac12  [\MM_{2}^{(-1)}, \widetilde {\MM}_{2}^{(-1)}],
\end{eqnarray}
where the first term with $\widehat \MM_3^{(-2)}$ is added, being a trivial solution to the above equation. Here 
$\widehat \MM_3^{(-2)}$ is an arbitrary trilinear map. 
This concludes the explicit computation of $\MM_3^{(-2)}$. 

Let us now move to the $2$-product $\MM_2^{(-2)}$. 
We have 
\begin{eqnarray}
\label{cosF}
\MM_2^{(-2)} = [Z_{D'}, [Z_D, \MM_2^{(0)}] = [Z_{D'}, [[\MM_1^{(0)}, -i\, \Theta(\iota_D)], \MM_2^{(0)}], 
\end{eqnarray}
where we have used the definition of the PCO $Z_D = [\MM_2^{(0)}, -i \Theta(\iota_D)]$. Now, by Jacobi identity 
and using the Leibniz rule $[\MM_1^{(0)}, \MM_2^{(0)}]=0$, we get 
\begin{eqnarray}
\label{cosG}
\MM_2^{(-2)} = [Z_{D'}, [\MM_1^{(0)}, [ -i \, \Theta(\iota_D), \MM_2^{(0)}]]] = 
[Z_{D'}, [\MM_1^{(0)}, \widetilde \MM_2^{(-1)}]]\,, 
\end{eqnarray}
where $\widetilde \MM_2^{(-1)}$ is defined as in (\ref{cosA}). Using again the Jacobi identity and 
that the PCO is closed, \emph{i.e.} $[\MM_1^{(0)}, Z_{D'}]=0$, this yields
\begin{eqnarray}
\label{cosH}
\MM_2^{(-2)} = [\MM_1^{(0)}, [Z_{D'}, \widetilde \MM_2^{(-1)}]]. 
\end{eqnarray}
Thus, we can define 
\begin{eqnarray}
\label{cosHA}
\widetilde \MM_2^{(-2)} = [Z_{D'}, \widetilde \MM_2^{(-1)}]\,. 
\end{eqnarray}
Inserting this result into the following defining equation for $\MM_3^{(-4)}$ 
\begin{eqnarray}
\label{cosI}
\frac12 [\MM_2^{(-2)} , \MM_2^{(-2)} ] + [ \MM_1^{(0)}, \MM_3^{(-4)} ]=0  
\end{eqnarray}
we finally get 
\begin{eqnarray}
\label{cosL}
 {\MM}_{3}^{(-4)} = [\MM_1^{(0)}, \widehat \MM_3^{(-4)}] 
 + \frac12  [\MM_{2}^{(-2)}, \widetilde {\MM}_{2}^{(-2)}]\,,
\end{eqnarray}
where $\widehat \MM_3^{(-4)}$ is again the trivial term, as above. In addition, from \eqref{cosHA}, we have 
\begin{eqnarray}
\label{cosM}
\widetilde \MM_2^{(-2)} = \Big[[\MM_1^{(0)}, - i \, \Theta(\iota_{D'})], \widetilde \MM_2^{(-1)}\Big]
= \Big[\MM_1^{(0)}, \widetilde{\widetilde \MM}_2^{(-2)}\Big] - [-i \, \Theta(\iota_{D'}), \MM^{(-1)}_2]
\,,
\end{eqnarray}
using again the Jacobi identities and where we put 
\begin{eqnarray}
\label{cosMA}
 \widetilde{\widetilde \MM}_2^{(-2)} = [- i\, \Theta(\iota_{D'}), \widetilde \MM_2^{(-1)}]. 
\end{eqnarray}
 As a consistency check, we have that
\begin{eqnarray}
\label{cosN}
 \Big[\MM_1^{(0)}, \widetilde \MM_2^{(-2)} \Big] = 
 -  \Big[\MM_1^{(0)},  [-i \, \Theta(\iota_{D'}), \MM^{(-1)}_2]\Big] = 
 [Z_{D'},  \MM^{(-1)}_2] = \MM_2^{(-2)}\,. 
\end{eqnarray}
which reproduces (\ref{cosA}) for the picture 2. 

\subsection{Some Explicit Examples of Computations}

In order to illustrate and make more clear the constructions in the previous section, we discuss specific examples 
of products of forms. We take into consideration the 2-products $M_2^{(-l)}$ with $l=0,1,2$ and 
for them we consider a collections of $0$-, $1$- and $2$-forms 
of the following types 
\begin{eqnarray}
\label{exaA}
&&\omega^{(0|0)}_A = A(x,\theta)\,, ~~~~~~
\omega^{(0|1)}_B = B(x,\theta) \delta(d\theta^1)\,, ~~~~ \nonumber \\
&&\omega^{(0|1)}_{B'} = B'(x,\theta) \delta(d\theta^2)\,, ~~~~~~
\omega^{(0|2)}_C = C(x,\theta) \delta(d\theta^1) \delta(d\theta^2)\,, ~~~~
\end{eqnarray}
where $A,B, B'$ and $C$ are supefields.\footnote{In the following, we write as an upperscript 
only the picture number  $\omega^{(0|a)} \rightarrow \omega^{(a)}$ with $a=0,1,2$.}
First of all, in the table~1 the picture numbers of the resulting forms are listed.   
\begin{table}[h]
\begin{center}
\begin{tabular}{|c|c|c|c|c|c|c|}
\hline
 ~~ & 0 $\times$ 0 & 0 $\times$ 1  & 1 $\times$ 1  & 0 $\times$ 2 & 1 $\times$ 2 & 2 $\times$ 2  \\
 \hline 
 $M_2^{(0)}$  & 0  & 1 & 2 & 2 & / & / \\
 $M_2^{(-1)}$ & / & 0 & 1 & 1 & 2 & / \\
$M_2^{(-2)}$  & / & / & 0  & 0 & 1 & 2 \\
\hline
\end{tabular}
\caption{In the table, we compute $M_2^{(-l)}(\omega^{(a)}, \omega^{(b)})$ where the pictures 
$a,b$ are listed in the first row as $a \times b$. The slanted line $/$ denotes a trivial result, 
while the numbers in the other  boxes denote the picture of the resulting form.}
\end{center}
\end{table}

\noindent 
Since $M_2^{(0)}$ is the usual wedge product we have 
\begin{eqnarray}
\label{exaB}
M_2^{(0)}(\omega^{(0)}_{A}, \omega^{(0)}_{A'}) &=& A A'\,, \nonumber \\
M_2^{(0)}(\omega^{(0)}_{A}, \omega^{(1)}_{B'}) &=& A B' \delta(d\theta^2) \,, \nonumber \\
M_2^{(0)}(\omega^{(1)}_{B}, \omega^{(1)}_{B'}) &=& B B' \delta(d\theta^1) \delta(d\theta^2) \,, \nonumber \\
M_2^{(0)}(\omega^{(0)}_{A}, \omega^{(2)}_{C}) &=& A C \delta(d\theta^1) \delta(d\theta^2) \,, \nonumber \\
M_2^{(0)}(\omega^{(1)}_{B}, \omega^{(2)}_{C}) &=& 0 \,, \nonumber \\
M_2^{(0)}(\omega^{(2)}_{C}, \omega^{(2)}_{C'}) &=& 0 \,, 
\end{eqnarray}
The last two expressions vanish since there is no picture 3 or 4 in our case. 
Let us compute now $M_2^{(-1)}$
\begin{eqnarray}
\label{exaC}
M_2^{(-1)}(\omega^{(0)}_{A}, \omega^{(0)}_{A'}) &=& 0 \,, \nonumber \\
M_2^{(-1)}(\omega^{(0)}_{A}, \omega^{(1)}_{B'}) &=&\frac13
\Big[Z_D\Big(A B' \delta(d\theta^2)\Big) + A Z_D \Big(B' \delta(d\theta^2)\Big) \Big] \,, \nonumber \\
M_2^{(-1)}(\omega^{(1)}_{B}, \omega^{(1)}_{B'}) &=& 
\frac13 \Big[
Z_D\Big(B B' \delta(d\theta^1) \delta(d\theta^2) \Big) 
+ Z_D\Big(B \delta(d\theta^1)\Big) B' \delta(d\theta^2) 
+ B \delta(d\theta^1) Z_D\Big(B' \delta(d\theta^2)\Big)  
\Big]
 \,, \nonumber \\
M_2^{(-1)}(\omega^{(0)}_{A}, \omega^{(2)}_{C}) &=& 
\frac13 \Big[ 
Z_D\Big( A C \delta(d\theta^1) \delta(d\theta^2)\Big) 
+ A Z_D\Big(C \delta(d\theta^1) \delta(d\theta^2)\Big) \Big]\,, \nonumber \\
M_2^{(-1)}(\omega^{(1)}_{B}, \omega^{(2)}_{C}) &=& 
\frac13 \Big[ 
Z_D \Big( B \delta(d\theta^1) \Big)  C \delta(d\theta^1) \delta(d\theta^2) + 
B \delta(d\theta^1) Z_D\Big(C \delta(d\theta^1) \delta(d\theta^2)  \Big) 
\Big]
\,, \nonumber \\
M_2^{(-1)}(\omega^{(2)}_{C}, \omega^{(2)}_{C'}) &=& 0 \,, 
\end{eqnarray}
Explicitly, we have 
\begin{eqnarray}
\label{exaD}
Z_D\Big(A B' \delta(d\theta^2)\Big) &=& d \Big[ -i \Theta(\iota_D) \Big(A B' \delta(d\theta^2)\Big) \Big] 
 -i \Theta(\iota_D) d \Big(A B' \delta(d\theta^2)\Big) \nonumber \\ 
 &=& \partial_2 (A B')\,, \nonumber \\
Z_D\Big(B' \delta(d\theta^2)\Big) &=& d \Big[-i \Theta(\iota_D) \Big(B' \delta(d\theta^2)\Big) \Big] -i \Theta(\iota_D) d \Big(B' \delta(d\theta^2)\Big) \nonumber \\
  &=& \partial_2 B'\,, \nonumber \\
 Z_D\Big(B \delta(d\theta^1)\Big) &=& d \Big[ -i\Theta(\iota_D) \Big(B \delta(d\theta^1)\Big) \Big]-i
 \Theta(\iota_D) d \Big(B \delta(d\theta^1)\Big) \nonumber \\
  &=& \partial_1 B\,, \nonumber \\
 Z_D\Big(A C \delta(d\theta^1) \delta(d\theta^2)\Big) &=& 
 d \Big[ -i\Theta(\iota_D) \Big(A C \delta(d\theta^1) \delta(d\theta^2)\Big) \Big] -i 
 \Theta(\iota_D) d \Big(A C \delta(d\theta^1) \delta(d\theta^2)\Big) \nonumber \\
 &=& - 2 D^\a \partial_\alpha (AC) \delta(D\cdot d\theta) \,, \nonumber 
\end{eqnarray}
\begin{eqnarray}
Z_D\Big(C \delta(d\theta^1) \delta(d\theta^2)\Big) &=& \nonumber
 d \Big[ -i\Theta(\iota_D) \Big(C \delta(d\theta^1) \delta(d\theta^2)\Big) \Big] -i 
 \Theta(\iota_D) d \Big(C \delta(d\theta^1) \delta(d\theta^2)\Big) \nonumber \\
 &=& - 2 D^\a \partial_\alpha C \delta(D\cdot d\theta) \,.
\end{eqnarray}
Finally
\begin{eqnarray}
\label{exaE}
M_2^{(-1)}(\omega^{(0)}_{A}, \omega^{(0)}_{A'}) &=& 0 \,, \nonumber \\
M_2^{(-1)}(\omega^{(0)}_{A}, \omega^{(1)}_{B'}) &=&\frac13 \Big(  \partial_2 (A B') + A \partial_2 B' 
\Big) 
\,, \nonumber \\
M_2^{(-1)}(\omega^{(1)}_{B}, \omega^{(1)}_{B'}) &=& 
\frac13 \Big[ 
- 2 D^\a \partial_\alpha (B B') \delta(D\cdot d\theta) + 
 \partial_1 B B' \delta(d\theta^2)  -  
 B  \partial_2 B'  \delta(d\theta^1)
\Big]
 \,, \nonumber \\
M_2^{(-1)}(\omega^{(0)}_{A}, \omega^{(2)}_{C}) &=& 
\frac13 \Big[ 
- 2 D^\a \partial_\alpha (AC) \delta(D\cdot d\theta)
 - 2 A D^\a \partial_\alpha C \delta(D\cdot d\theta) \Big] \nonumber \\
 &=& 
- \frac{2}{3} \Big[ 
D^\a \partial_\alpha (AC)
 +A D^\a \partial_\alpha C \Big] \delta(D\cdot d\theta) \,, \nonumber \\
M_2^{(-1)}(\omega^{(1)}_{B}, \omega^{(2)}_{C}) &=& 
\frac13 \Big[ 
\partial_1 B C \delta(d\theta^1) \delta(d\theta^2) + 2 B D^\a \partial_\a C  \delta(d\theta^1) \delta(D\cdot d\theta)
\Big]
 \nonumber \\
&=& \frac13 \Big[ 
\partial_1 B C + \frac{2}{D^1} B D^\a \partial_\a C 
\Big] \delta(d\theta^1) \delta(d\theta^2)\,, \nonumber \\
M_2^{(-1)}(\omega^{(2)}_{C}, \omega^{(2)}_{C'}) &=& 0 \,, 
\end{eqnarray}
The resulting products have the correct picture assignment as depicted in the table 1. 
We notice that the result depends upon the choice of the PCO and therefore they depend 
upon the odd vector field $D$. \\
Now, we consider the last case: $M_2^{(-2)}$. We have 
\begin{eqnarray}
\label{exaF}
M_2^{(-2)}(\omega^{(0)}_{A}, \omega^{(0)}_{A'}) &=& 0 \,, \nonumber \\
M_2^{(-2)}(\omega^{(0)}_{A}, \omega^{(1)}_{B'}) &=&0 
 \,, \nonumber \\
M_2^{(-2)}(\omega^{(1)}_{B}, \omega^{(1)}_{B'}) &=& \frac19\Big[
Z_{D'} Z_D
\Big(B B' \delta(d\theta^1) \delta(d\theta^2) 
\Big) + 
Z_{D'} \Big( M_2^{(-1)}(\omega^{(1)}_{B}, \omega^{(1)}_{B'})\Big)
 + 
Z_{D} \Big( M_2^{\prime(-1)}(\omega^{(1)}_{B}, \omega^{(1)}_{B'}) 
\Big)
\nonumber \\
&+& 
Z_{D'}\Big(B \delta(d\theta^1) \Big) Z_{D}\Big(B' \delta(d\theta^2) \Big) + 
Z_{D}\Big(B \delta(d\theta^1) \Big) Z_{D'}\Big(B' \delta(d\theta^2 \Big) \Big]\,, 
\nonumber \\
M_2^{(-2)}(\omega^{(0)}_{A}, \omega^{(2)}_{C}) &=& 
\frac19 \Big[ 
 Z_{D'} Z_D \Big( A C \delta(d\theta^1) \delta(d\theta^2) \Big) + 
Z_{D'} M_2^{(-1)}(\omega_A^{(0)},\omega_C^{(2)}) + 
Z_{D} M_2^{\prime (-1)}(\omega_A^{(0)},\omega_C^{(2)}) \nonumber \\
&+& 
A Z_{D'} 
 Z_D\Big(C \delta(d\theta^1) \delta(d\theta^2)\Big) \Big]\,, 
 \nonumber \\
M_2^{(-2)}(\omega^{(1)}_{B}, \omega^{(2)}_{C}) &=& 
\frac19 \Big[ 
Z_{D'} M_2^{(-1)}(\omega_B^{(1)},\omega_C^{(2)}) + 
Z_{D} M_2^{\prime (-1)}(\omega_B^{(1)},\omega_C^{(2)}) \nonumber \\
&+& 
Z_{D'}\Big(  B\delta(d\theta^1)\Big) Z_D\Big(C \delta(d\theta^1) \delta(d\theta^2)\Big)
+ 
Z_{D}\Big(  B\delta(d\theta^1)\Big) Z_{D'}\Big(C \delta(d\theta^1) \delta(d\theta^2)\Big)
\nonumber \\
&+&
B\delta(d\theta^1) 
 Z_{D'} 
 Z_D\Big(C \delta(d\theta^1) \delta(d\theta^2)\Big) \Big]\,, 
 \nonumber \\
M_2^{(-2)}(\omega^{(2)}_{C}, \omega^{(2)}_{C'}) &=& 
\frac19 \Big[ 
Z_{D'} Z_D \Big(C \delta(d\theta^1) \delta(d\theta^2)\Big)
 \Big(C' \delta(d\theta^1) \delta(d\theta^2)\Big) 
 \nonumber \\
&+&
Z_{D'}\Big(C \delta(d\theta^1) \delta(d\theta^2)\Big) Z_D\Big(C' \delta(d\theta^1) \delta(d\theta^2)\Big)
\nonumber \\
&+&
Z_{D}\Big(C \delta(d\theta^1) \delta(d\theta^2)\Big) Z_{D'}\Big(C' \delta(d\theta^1) \delta(d\theta^2)\Big)
  \nonumber \\
&+&
 \Big(C \delta(d\theta^1) \delta(d\theta^2)\Big)  Z_{D'} Z_D \Big(C' \delta(d\theta^1) \delta(d\theta^2)\Big)\Big]
 \,. 
\end{eqnarray}
where we have denoted 
$M_2^{\prime (-1)}$ the 2-product with respect to the odd vector $D'$. 
Each single piece is computed as follows
\begin{eqnarray}
\label{exaG}
Z_{D'} \Big( M_2^{(-1)}(\omega^{(1)}_{B}, \omega^{(1)}_{B'})\Big) &=& 
- \frac{2}{ 3}
\epsilon^{\b\a}\partial_\b\partial_\alpha (B B') 
+ \frac{1}{ 3 D^{\prime 1}}D^{\prime \b} \partial_\b (\partial_1 B B') 
- \frac{1}{ 3 D^{\prime 2}}D^{\prime \b} \partial_\b (B  \partial_2 B' )  \,,
\nonumber \\
Z_{D}  \Big( M_2^{\prime (-1)}(\omega_B^{(1)},\omega^{(1)}_{B'})  \Big)&=& 
- \frac{2}{ 3}
\epsilon^{\b\a}\partial_\b\partial_\alpha (B B') 
+ \frac{1}{ 3 D^{1}}D^{\b} \partial_\b (\partial_1 B B') 
- \frac{1}{ 3 D^{2}}D^{\b} \partial_\b (B  \partial_2 B' ) \,,
\nonumber \\
Z_{D}\Big(  B\delta(d\theta^1)\Big) &=& \partial_1 B\,, \nonumber \\
Z_{D'}\Big(  B\delta(d\theta^1)\Big) &=& \partial_1 B\,, \nonumber \\
Z_{D}\Big(  B'\delta(d\theta^2)\Big) &=& \partial_2 B' \,, \nonumber \\
Z_{D'}\Big(  B' \delta(d\theta^2)\Big) &=& \partial_2 B' \,, \nonumber \\
Z_{D'}  \Big( M_2^{(-1)}(\omega_B^{(1)},\omega_C^{(2)})  \Big) &=& 
 - \frac23 D^{\prime \a}\partial_\a \Big[ 
\partial_1 B C + \frac{2}{D^1} B D^\a \partial_\a C 
\Big] \delta(D'\cdot d\theta)\,,
\nonumber 
\end{eqnarray}
\begin{eqnarray}
Z_{D}  \Big( M_2^{\prime(-1)}(\omega_B^{(1)},\omega_C^{(2)})  \Big)  &=& 
 - \frac23 D^{\a}\partial_\a \Big[ 
\partial_1 B C + \frac{2}{D^{\prime 1}} B D^{\prime \a} \partial_\a C 
\Big] \delta(D\cdot d\theta)\,,
 \nonumber \\
Z_D\Big(C \delta(d\theta^1) \delta(d\theta^2) \Big) &=& - 2 D^\a \partial_\a C \delta(D\cdot d\theta)\,, \nonumber \\
Z_{D'}\Big(C \delta(d\theta^1) \delta(d\theta^2) \Big) &=& 
- 2 D^{\prime \a} \partial_\a C \delta(D'\cdot d\theta)\,,
\nonumber \\
Z_{D'} Z_D \Big(B B' \delta(d\theta^1) \delta(d\theta^2) \Big) &=& 
2 \epsilon^{\a\b} \partial_\a \partial_\b (B B')\,, \nonumber \\
Z_D Z_{D'} \Big(C \delta(d\theta^1) \delta(d\theta^2) \Big) &=& 2 \epsilon^{\a\b} \partial_\a \partial_\b C\,.
\end{eqnarray}
Now, it is a simple matter to replace each single pieces into the definitions given in the previous equations. 
For example, we get 
\begin{eqnarray}
\label{exaH}
M_2^{(-2)}(\omega^{(2)}_{C}, \omega^{(2)}_{C'}) &=& \frac19 
\Big[ 
(\epsilon^{\a\b} \partial_\a \partial_\b C) C' \delta(d\theta^1) \delta(d\theta^2) 
+ 4 D^{\prime \a} \partial_\a C \delta(D'\cdot d\theta) 
D^\a \partial_\a C' \delta(D\cdot d\theta) \nonumber \\
&+& 4 D^\a \partial_\a C \delta(D\cdot d\theta) 
D^{\prime \a} \partial_\a C' \delta(D'\cdot d\theta) + 
 C \delta(d\theta^1) \delta(d\theta^2) (\epsilon^{\a\b} \partial_\a \partial_\b C')
\Big]  \nonumber \\
&=& 
\frac19 
\Big[ 
(\epsilon^{\a\b} \partial_\a \partial_\b C) C' 
+ 2 \epsilon^{\a\b} \partial_\a C \partial_\b C' + 
 C (\epsilon^{\a\b} \partial_\a \partial_\b C')
\Big] \delta(d\theta^1) \delta(d\theta^2) \nonumber \\
&=&\frac19 \epsilon^{\a\b} \partial_\a \partial_\b (C C')  \delta(d\theta^1) \delta(d\theta^2)\,. 
\end{eqnarray}
which is independent of $D$ and of $D'$. Since $M_2^{(-2)}$ maps two $2$-picture forms into a $2$-picture form
it also preserves the invariance under $SL(2)$ isometries and therefore the result can be written in 
manifestly invariant way. 

In the same way, one can compute the other expressions. Finally, we can check the non-associativity 
for the last expression, namely we can check that
\begin{eqnarray}
\label{exaI}
M_2^{(-2)}\Big(
\omega^{(2)}_{C}, M_2^{(-2)}(\omega^{(2)}_{C'},  \omega^{(2)}_{C''}) \Big) 
+ 
M_2^{(-2)}\Big(
M_2^{(-2)}(\omega^{(2)}_{C}, \omega^{(2)}_{C'}),  \omega^{(2)}_{C''} \Big) 
\neq 0\,. 
\end{eqnarray}
leading to a $3$-product $M_3^{(-4)}$ source of the $A_\infty$-algebra. 


\section*{Acknowledgements}

We would like to thank R. Donagi, C. Maccaferri, I. Sachs for very useful discussions.

\setcounter{equation}{0}
\renewcommand{\theequation}{A.\arabic{equation}}

\section*{Appendix A: How to compute with $\Theta(\iota_D)$ and $\delta(\iota_D)$} 

In  order to clarify the action of $\Theta(\iota_D), \delta(\iota_D)$ and $Z_D$,
we present  some detailed calculations.
Let us compute the action of $\Theta(\iota_D)$ on $\delta(d\theta^\a)$ with $\a =1,2$.
\begin{eqnarray}
\label{ant_AB}
\Theta(\iota_D) \delta(d\theta^\a) &=&  - i \lim_{\epsilon \rightarrow 0 }\int_{-\infty}^\infty dt \frac{e^{i t \iota_D}}{t + i \epsilon} \delta(d\theta^\a)
= -i \lim_{\epsilon \rightarrow 0 } \int_{-\infty}^\infty dt  \frac{ \delta(d\theta^\a + i D^\a t)}{t + i \epsilon} \nonumber \\
&=& -
\frac{1}{D^\a}  \lim_{\epsilon \rightarrow 0 } \int_{-\infty}^\infty dt \frac{ \delta(t - i \frac{d\theta^\a}{D^\a})}{t + i \epsilon} =
\frac{i}{d\theta^\a} \in \Omega^{(-1|0)}_{\proj{1|2}}
\end{eqnarray}
where the coefficient $D^\a$ drops out from the computation (but it must be different from zero in order to have a meaningful computation). In the same way, we have
\begin{eqnarray}
\label{ant_AC}
\delta(\iota_D) \delta(d\theta^\a) =  \int_{-\infty}^\infty dt {e^{i t \iota_D}} \delta(d\theta^\a)  =
  \int_{-\infty}^\infty dt \delta(d\theta^\a + i D^\a t) =
 -\frac{i}{ D^\a} \in \Omega^{(0|0)}_{\proj {1|2}}\,,
\end{eqnarray}
using the distributional properties. Again the requirement that $D^\a$ is different from zero is crucial.

Let us compute the action of $\Theta(\iota_D)$ on the
product $d\theta^\b \delta(d\theta^\a)$. We assume that $\a\neq \b$, otherwise it vanishes. Applying the
same rules we have
\begin{eqnarray}
\label{ant_ACA}
\Theta(\iota_D) \Big( d\theta^\b \delta(d\theta^\a) \Big) &=&
- i  \lim_{\epsilon \rightarrow 0 } \int_{-\infty}^\infty dt \frac{e^{i t \iota_D}}{t + i \epsilon}\Big( d\theta^\b
\delta(d\theta^\a) \Big) \nonumber \\
&= &  -i  \lim_{\epsilon \rightarrow 0 } \int_{-\infty}^\infty dt  \frac{  (d\theta_\b + i D_\b t) \delta(d\theta^\a + i D^\a t)}{t + i \epsilon} \nonumber \\
 & = &
  \frac{-i}{i D^\a} \lim_{\epsilon \rightarrow 0 } \int_{-\infty}^\infty dt  \frac{  (d\theta^\b + i D^\b t)}{t + i \epsilon} \delta
  \Big(t - \frac{i d\theta_\a }{D^\a}\Big) \nonumber \\
  & = &
  -\frac{1}{D^\a} \Big( d\theta_\b + i D^\b \frac{i d\theta^\a}{D_\a}\Big) \frac{D^\a}{i d\theta^\a} \nonumber \\
  &=& i  \Big( \frac{d\theta^\b}{d\theta^\a} -  \frac{D^\b}{D^\a}\Big) \in \Omega^{(0|0)}_{\proj {1|2}}
\end{eqnarray}
from which it immediately follows that if $\alpha = \beta$, then both members vanish.
Analogously, we have 
\begin{eqnarray}
\label{ant_ACAB}
\delta(\iota_D) \Big( d\theta^\b \delta(d\theta^\a) \Big) &=&
\int_{-\infty}^\infty dt e^{i t \iota_D} \Big( d\theta^\b
\delta(d\theta^\a) \Big) =
  \int_{-\infty}^\infty dt   (d\theta_\b + i D_\b t) \delta(d\theta^\a + i D^\a t) \nonumber \\
 &=&
  \frac{1}{i D^\a}  \int_{-\infty}^\infty dt   (d\theta^\b + i D^\b t) 
  \delta \Big(t - \frac{i d\theta_\a }{D^\a}\Big) \nonumber \\
  &=& \frac{1}{i D^\a} \Big( d\theta^\b  -  \frac{D^\a}{D^\b}  d\theta^\a \Big) \in \Omega^{(1|0)}_{\proj {1|2}}
\end{eqnarray}
which also vanishes if $\alpha =\beta$. 

Let us also consider the following expressions
\begin{eqnarray}
\label{ant_ACB}
\Theta(\iota_D) \Big( \frac{1}{d\theta_\b} \delta(d\theta_\a) \Big) &=&
 -i \lim_{\epsilon \rightarrow 0 }  \int_{-\infty}^\infty dt  \frac{  \delta(d\theta_\a + i D_\a t)}{(d\theta_\b + i D_\b t)
 (t + i \epsilon)} \nonumber \\
 &=&
  \frac{-i}{i D_\a} \lim_{\epsilon \rightarrow 0 } \int_{-\infty}^\infty dt  \frac{1}{(d\theta_\b + i D_\b t)(t + i \epsilon)} \delta
  \Big(t - \frac{i d\theta_\a }{D_\a}\Big) \nonumber \\
  &=&
  -\frac{1}{D_\a} \frac{1}{\Big( d\theta_\b + i D_\b \frac{i d\theta_\a}{D_\a}\Big)} \frac{D_\a}{i d\theta_\a} \nonumber \\
  &=& i  \frac{1}{\Big( \frac{d\theta_\b}{d\theta_\a} -  \frac{D_\b}{D_\a}\Big)} \frac{1}{d\theta_\a^2} \in \Omega^{(-2|0)}_{\proj {1|2}}
\end{eqnarray}
which is an inverse form. Notice that if $\alpha=\beta$, the product
$ \Big( \frac{1}{d\theta^\b} \delta(d\theta^\a) \Big) $ is ill-defined, and this is consistent
with the fact that also the right-hand side is divergent.

Let us now compute the action of $\Theta(\iota_D)$ on $\Omega^{(0|2)}_{\proj {1|2}}$.
This is done as follows
\begin{eqnarray}
\label{ant_AD}
\Theta(\iota_D) \Big( \delta(d\theta_1) \delta(d\theta_2) \Big) &=&
-i \lim_{\epsilon \rightarrow 0 } \int_{-\infty}^\infty dt \frac{e^{i t \iota_D}}{t + i \epsilon} \delta(d\theta_1)  \delta(d\theta_2) \nonumber \\
&=& -i
 \lim_{\epsilon \rightarrow 0 }\int_{-\infty}^\infty dt \frac{\delta(d\theta_1 + i t D_1)  \delta(d\theta_2 + i t D_2)}{t + i \epsilon} \nonumber \\
 &=&  \frac{i}{D_1 D_2} \left( D_1
 \frac{\delta\Big(d\theta_2 - \frac{D_2}{D_1} d\theta_1\Big)}{d\theta_1}
 - D_2  \frac{\delta\Big(d\theta_1 - \frac{D_1}{D_2} d\theta_2\Big)}{d\theta_2} \right) \nonumber\\
 &=& - \frac{i}{D_1 D_2} \Big(\frac{D_1}{d\theta_1} + \frac{D_2}{d\theta_2} \Big)  \delta\Big(
 \frac{d\theta_1}{D_1}  - \frac{d\theta_2}{D_2} \Big) \nonumber \\
 &=& = -i  \Big(\frac{D_1}{d\theta_1} + \frac{D_2}{d\theta_2} \Big)  \delta(D\cdot d\theta) \in \Omega^{(-1|1)}_{\proj {1|2}}\,.
\end{eqnarray}
where $(D \cdot d\theta) = D_\a \epsilon^{\a\b} d\theta_\b$.

Notice that the linear combination of $d\theta_1$ and $d\theta_2$
appearing in the first factor is linearly independent from the linear combination appearing in the Dirac delta
argument. Notice also that the sign between the two Dirac delta's in the second line is due to
the fermionic nature of $dt$ and of the Dirac delta form. This
sign is crucial for the left-hand side and the right-hand side of eq.~(\ref{ant_AD}) be consistent.
Indeed, if we interchange $\delta(d\theta_1)$ with $\delta(d\theta_2)$ in the left-hand side we get an overall minus sign; on the other hand, on the right-hand side of the equation,
by exchanging $d\theta_1$ and $d\theta_2$ in the Dirac delta argument again a sign emerges.

Finally, we can consider another independent odd vector field $D'$ and the corresponding operator $\Theta(\iota_{D'})$.
Acting on (\ref{ant_AD}) it yields
\begin{eqnarray}
\label{act_ADA}
\Theta(\iota_{D'}) \Theta(\iota_D) \Big( \delta(d\theta_1) \delta(d\theta_2) \Big) =
\frac{\det(D', D)}{(D' \cdot d\theta) (D \cdot d\theta)} \in \Omega^{(-2|0)}_{\proj {1|2}}
\end{eqnarray}
 where $(D \cdot d\theta) = D_\a \epsilon^{\a\b} d\theta_\b$ and
 $\det(D',D) = D'_\a \epsilon^{\a\b} D_\b = D' \cdot D$.
Notice that in this case, by interchanging $\delta(d\theta_1)$ with
$\delta(d\theta_2)$, we get again an overall minus sign. This is
obtained also by exchanging the coefficients of the vectors $D$ and $D'$,
and in this way we get a minus sign from the determinant $\det(D', D)$.

Let us also consider the action of $\delta(\iota_D)$ on the product of $\delta(d\theta^1)\delta(d\theta^2)$. We have 
\begin{eqnarray}
\label{new_ACT}
\delta(\iota_D) (\delta(d\theta^1)\delta(d\theta^2)) = - i \delta( D\cdot d\theta) \in \Omega^{(0|1)}_{\proj {1|2}}, 
\end{eqnarray}
and finally 
\begin{eqnarray}
\label{new_ACT}
\delta(\iota_{D^\prime}) 
\delta(\iota_D)( \delta(d\theta^1)\delta(d\theta^2) ) = {\rm det}(D', D) \in \Omega^{(0|0)}_{\proj {1|2}}, 
\end{eqnarray}
which also follows from (\ref{act_ADA}) by the identity $ d\theta^\a \Theta(\iota_{D}) = \delta(d\theta^\a)$. 

The action of a second PCO decreases the picture number as to bring elements of $\Omega^{p|2}_{\mathbb{P}^{1|2}}$ into superforms having picture number equal to zero.
Note that since the PCO's $Z$ is formally exact as stress above,
it maps cohomology classes into cohomology classes, $H_{dR}^{(p|2)} \rightarrow H_{dR}^{(p|0)}$, therefore
it is natural to expect that it can only properly acts on cohomology classes, and indeed, acting on representatives of $H_{dR}^{(p|2)}$ one never gets inverse forms.
Nonetheless, it can be shown that acting on generic elements of the space $\Omega^{(p|2)}_{\mathbb{P}^{1|2}}$, not necessarily closed,
one never produces inverse forms. Let us show this first in a very simple example.

\noindent Consider a generic integral form in $\Omega^{1|2}_{\mathbb{P}^{1|2}} \cong \mathcal{B}er (\proj {1|2})$
\begin{eqnarray}
\label{ant_B}
\omega^{(1|2)}= A(z,\theta) dz \delta(d\theta^1) \delta(d\theta^2)
\end{eqnarray}
where $A(z,\theta^\alpha)$ is a superfield in the local coordinates of $\mathbb{P}^{1|2}$. Being (the analog of) a top-form
it is naturally closed. Acting with $Z_D$ one gets
\begin{eqnarray}
\label{ant_C}
Z_D(\omega^{(1|2)}) &=& d\left[ -i\Theta(\iota_D) A dz \delta(d\theta^1) \delta(d\theta^2)\right] -i
 \Theta(\iota_D) \left[ d\left( A dz \delta(d\theta^1) \delta(d\theta^2) \right)\right]\nonumber \\
 &=& d \Big[  A \, \Big( \frac{D^1}{d\theta^1} + \frac{D^2}{d\theta^2} \Big) dz \, \delta(D\cdot d\theta) \Big] \nonumber \\
 &=&
  2 \Big( (D^1 \partial_1 A  + D^2 \partial_2 A) \, dz\, \delta(D\cdot d\theta) \Big) \nonumber \\
 &=&
  2 D^\alpha \partial_\alpha A \, dz\, \delta(D\cdot d\theta) \Big) \in \Omega^{(1|1)}_{\proj {1|2}},
  \end{eqnarray}
where $\partial_\alpha A$ are the derivatives with respect to $\theta^\alpha$ of the superfield $A$.
The result is in $\Omega^{(1|1)}_{\mathbb{P}^{1|2}}$, it is closed and no inverse form is required.
However, the form (\ref{ant_C}) is not the most general $(1|1)$-pseudoform. 

\noindent Let us act with a second PCO :
\begin{eqnarray}
\label{ant_D}
Z_{D'}\Big[ 2  D^\alpha \partial_\alpha A  \, dz\, \delta(D\cdot d\theta) \Big) \Big]
&=& 2 \epsilon^{\a\b} \partial_\a \partial_\b A \, dz \in \Omega^{(1|0)}_{\proj {1|2}}
\end{eqnarray}
which is a superform in $\Omega^{(1|0)}_{\mathbb{P}^{1|2}}$, it does not contain
any inverse form and it is independent of the  odd vector fields $D, D'$.   Note that this particular
expression is closed, since $\partial_1^2 = \partial^2_2 = \{\partial_1, \partial_2\}=0$.
No inverse form is needed in the present case.

\newcommand{\sdot}{\,\mbox{\tiny $\bullet$}\,}


\bibliographystyle{amsplain}

\begin{thebibliography}{9}


\bibitem{Aspinwall} P. S. Aspinwall {\it et al}, \emph{Dirichlet Branes and Mirror Symmetry}, Chapter 8, Clay Mathematics Monographs (Vol 4)

\bibitem{Belo1} A. Belopolsky, \emph{Picture Changing Operators in Supergeometry and Superstring Theory}, arXiv:9706033 [hep-th]

\bibitem{Berkovits:2004px}
  N.~Berkovits,
  \emph{Multiloop amplitudes and vanishing theorems using the pure spinor formalism for the superstring,}
  JHEP {\bf 0409} (2004) 047
  [hep-th/0406055].

\bibitem{Berkovits:2004dt}
  N.~Berkovits,
  \emph{Covariant multiloop superstring amplitudes,}
  Comptes Rendus Physique {\bf 6} (2005) 185
  [hep-th/0410079].

\bibitem{Berkovits:2006vi}
  N.~Berkovits and N.~Nekrasov,
  \emph{Multiloop superstring amplitudes from non-minimal pure spinor formalism,}
  JHEP {\bf 0612} (2006) 029
  doi:10.1088/1126-6708/2006/12/029
  [hep-th/0609012].

\bibitem{Berkovits:1994vy} 
  N.~Berkovits and C.~Vafa,
  \emph{$N=4$ topological strings},
  Nucl.\ Phys.\ B {\bf 433}, 123 (1995)

  

  








\bibitem{CN} S.L. Cacciatori, S. Noja, \emph{Projective Superspaces in Practice}, J. Geom. Phys. {\bf 130}, 40-62 (2018)

\bibitem{CNR} S.L. Cacciatori, S. Noja, R. Re, \emph{Non Projected Calabi-Yau Supermanifolds over $\mathbb{P}^2$}, arXiv:1706.01354 [math.AG]

\bibitem{CCGhd} L. Castellani, R. Catenacci, P.A. Grassi, \emph{Hodge Dualities on Supermanifolds}, Nucl. Phys. B {\bf 899}, 570 (2015)

\bibitem{CCGir} L. Castellani, R. Catenacci, P.A. Grassi, \emph{Integral representations on supermanifolds: super Hodge duals, PCOs and Liouville forms}, Lett. Math. Phys, {\bf 107}, 1, 167-180 (2017)



\bibitem{CCGGeom} L. Castellani, R. Catenacci, P.A. Grassi, \emph{The Geometry of Supermanifolds and New Supersymmetric Actions}, Nucl. Phys. B {\bf 899}, 112 (2015)

\bibitem{CDGM} R. Catenacci, M. Debernardi, P.A. Grassi, D. Matessi, \emph{\v{C}ech and de Rham Cohomolgy of Integral Forms}, J. Geom. Phys. {\bf 62}, 890 - 902 (2012)

\bibitem{Catenacci:2018xsv} 
  R.~Catenacci, P.~A.~Grassi and S.~Noja, \emph{Superstring Field Theory, Superforms and Supergeometry},
  arXiv:1807.09563 [hep-th].



\bibitem{DonWit} R. Donagi, E. Witten, \emph{Supermoduli Space is Not Projected} Proc. Symp. Pure Math. {\bf 90} 19-72 (2015)

\bibitem{Erler} 
  T.~Erler, S.~Konopka and I.~Sachs,
 \emph{Resolving Witten`s superstring field theory},
  JHEP {\bf 1404}, 150 (2014)


\bibitem{FMS} D. Friedan, S. Shenker, E. Martinec, \emph{Conformal Invariance, Supersymmetry and String Theory}, Nucl. Phys. B {\bf 271} 93-165 (1986)

\bibitem{GWS} M.B. Green, J.H. Schwarz, E. Witten, \emph{Superstring Theory}, Vol. 1-2, CUP (1988)


\bibitem{Keller} B. Keller, \emph{Introduction to $A$-infinity Algebras and Modules}, Homology Homopoty Appl. {\bf 3}, 1, (2001) 1-35



\bibitem{Manin} Yu.I. Manin, \emph{Gauge Fields and Complex Geometry}, {%
Springer-Verlag} (1988)



\bibitem{1DCY} S. Noja, S.L. Cacciatori, F. Dalla Piazza, A. Marrani, R. Re, \emph{One-Dimensional Super Calabi-Yau Manifolds and their Mirrors}, JHEP {\bf 1704}, 094 (2017)


\bibitem{PiGeo} S. Noja, \emph{Supergeometry of $\Pi$-Projective Spaces}, J. Geom. Phys. {\bf 124}, 286-299 (2018)

\bibitem{Preitschopf:1989fc}
  C.~R.~Preitschopf, C.~B.~Thorn and S.~A.~Yost,
  \emph{Superstring Field Theory,}
  Nucl.\ Phys.\ B {\bf 337} (1990) 363.



\bibitem{pol} 
J. Polchinski, \emph{String Theory} Vol. 1-2, CUP (1998)


\bibitem{VorGeom} Th. Th. Voronov, \emph{Geometric Integration Theory on Supermanifolds}, Soviet Scientific Review, Section C: Mathematical Physics, {\bf 9}, Part 1, Harwood Academic Publisher (1992). Second Edition: Cambridge Scientific Publisher (2014)



\bibitem{Witten:1986qs}
  E.~Witten,
  \emph{Interacting Field Theory of Open Superstrings,}
  Nucl.\ Phys.\ B {\bf 276} (1986) 291.


\bibitem{Witten} E. Witten, \emph{Notes on Supermanifolds and Integrations}, arXiv:1209.2199 [hep-th]
  
\bibitem{WittenPerturbation}
  E.~Witten,
  \emph{Superstring Perturbation Theory Revisited},
  arXiv:1209.5461 [hep-th].







\end{thebibliography}

\end{document}